\documentclass{article}

\usepackage[final]{neurips_2026_vericode}

\usepackage[utf8]{inputenc}
\usepackage[T1]{fontenc}
\usepackage{hyperref}
\usepackage{url}
\usepackage{booktabs}
\usepackage{amsfonts}
\usepackage{nicefrac}
\usepackage{microtype}
\usepackage{xcolor}

\usepackage{graphicx}
\usepackage{amsmath,amssymb}
\usepackage{tabularx}
\usepackage{array}
\usepackage{xspace}

\setcitestyle{numbers,square}

\workshoptitle{AI for Verifiable Coding: Human-Aligned Collaborative Agents for Autoformalization, Proofs, and Heuristics}

\newcolumntype{Y}{>{\raggedright\arraybackslash}X}

\newcommand{\mergegym}{\textsc{MergeGym}\xspace}
\newcommand{\touch}{\textsc{OneSidedTouch}\xspace}
\newcommand{\linmodel}{\textsc{Lineage-LR}\xspace}
\newcommand{\mergecheck}{\texttt{git merge-tree}\xspace}

\title{
Asymmetric Repository Lineage Modeling and
Verifier-Guided Coordination in Concurrent AI Coding Agents
}

\author{%
  Arjun Subramanian\\
  MIT CSAIL\\
  \texttt{arjun\_s@mit.edu}
  \And
  George Xu\\
  Harvard Medical School\\
  \texttt{gexu@mgh.harvard.edu}
  \And
  Nithilan Karthik\\
  Managent\\
  \texttt{nithilan@managent.tech}
}
\begin{document}

\maketitle

\begin{abstract}
Concurrent AI coding agents create a coordination problem in which cheap signals may prioritize work, but only an executable checker can establish the property being claimed. We study this separation through a property-scoped verification contract and \mergegym, a three-track benchmark for open-time scope forecasting, replay-conditioned conflict resolution, and scheduling. On a stratified 715-pair lineage set (167 textual conflicts), 79 conflicts (47.3\%) occur despite disjoint authored file sets. This is an operationally important \emph{proxy mismatch expected from three-way merge}: authored PR diffs are measured against PR-specific bases, whereas the checker compares both heads to their common merge base. A standalone lineage-union rule reaches held-out AUROC 0.877; a 12-feature logistic model reaches 0.882 [0.845, 0.917] with PR-AUC 0.661, and an untuned random forest on the same decision-time features reaches 0.902 [0.875, 0.928] with PR-AUC 0.701. At a 33.3\% held-out replay budget, the logistic and random-forest models recover 81.2\% and 82.9\% of conflicts, respectively. Patch reconstruction succeeds for 48/79 zero-overlap conflicts and all 48 become clean; the other 31 cases are inconclusive, so this check validates the expected three-way-merge explanation rather than claiming a new Git mechanism. In T1, a zero-shot LLM reaches AUC 0.704 and LLM-plus-metadata fusion 0.740. In T3, a decision-time gate de-overlaps a median 91.7\% of labeled scope collisions at 65.0\% makespan inflation under frozen-label replay. Because local \mergecheck replay is already cheap in our logs (median 0.02 s), we do not claim that lineage triage saves this checker alone: when exact replay is cheap, verify everything. All empirical guarantees in this paper remain limited to textual mergeability or the explicitly stated frozen-label scheduling target.
\end{abstract}

\section{Introduction}
Modern coding agents can open, modify, test, and submit pull requests without waiting for a human developer. Once several agents operate on the same repository, reliability becomes a property of \emph{interacting work}, not merely of individually plausible patches. With $k$ concurrent PRs, a coordinator may face $k(k-1)/2$ pairwise interactions before considering higher-order effects. Two PRs can edit the same file, duplicate the same task in different files, silently invalidate one another through branch history, or merge textually while breaking build or semantic invariants.

This is naturally a verifier-in-the-loop problem. A learned score is useful only if its role is separated from the property it cannot certify. In our setting, \mergecheck deterministically evaluates one narrow predicate: whether two fixed Git heads admit a textual three-way merge from their common base. It does \emph{not} certify compilation, tests, semantic equivalence, security, or user intent. That narrowness is useful because it gives us a clean trusted boundary: prediction may decide \emph{what to check first}; the checker decides the textual property itself.

We study one systems question across three empirical scales: \emph{how should concurrent coding-agent workflows separate cheap coordination signals from authoritative checks?} Population measurements establish that co-active agent work is already common. \mergegym then decomposes pairwise coordination into stage-specific tasks instead of collapsing scope, merge outcome, resolution quality, and scheduling into one generic ``conflict'' label. Finally, a replay-labeled lineage study quantifies a specific proxy mismatch: authored-file overlap is computed from each PR against its own base, whereas a head-to-head three-way merge is defined relative to the two-head common base. Branch history can therefore make two disjoint authored diffs conflict. This behavior follows from Git semantics; our contribution is to measure how often it matters in agent-PR coordination and how much decision-time lineage state improves held-out ranking.

The paper makes four contributions:
\begin{enumerate}
  \item \textbf{A property-scoped verification contract.} We explicitly separate observable proxies, learned risk scores, and executable checks, and require every claim to remain attached to the backend that establishes it.
  \item \textbf{A staged benchmark for interacting agents.} \mergegym separates open-time scope forecasting, replay-conditioned conflict resolution, and online scheduling over 577,045 co-active pairs and 97 repository streams; the evaluation includes the stronger T1 LLM baselines present in the released artifact.
  \item \textbf{A quantified authored-diff proxy mismatch.} On 715 stratified replay-labeled pairs, 47.3\% of textual conflicts have disjoint authored file sets. A simple lineage-union rule reaches held-out AUROC 0.877, the transparent 12-feature logistic model reaches 0.882, and an untuned random forest using the same decision-time features reaches 0.902.
  \item \textbf{Decision-focused robustness.} We add held-out PR-AUC and replay-budget utility, feature-subset ablations, stronger tabular baselines, 30 alternative repository-held-out splits, agent/language group analyses, lineage-truncation sensitivity, and episode-level uncertainty for T3, while retaining exact replay as the authority whenever the textual property must be guaranteed.
\end{enumerate}

\section{Verification contract: separate proxy, predictor, and checker}
Let a co-active PR pair be $q=(A,B)$ with heads $h_A,h_B$ and common merge base $m$. We distinguish three kinds of signals.

\paragraph{Observable scope.}
Let $D_A,D_B$ be the final authored changed-file sets. The relation
\begin{equation}
O(q)=\mathbf{1}[D_A\cap D_B\neq\varnothing]
\end{equation}
is exact for authored file overlap, but it is not a correctness or conflict certificate.

\paragraph{Executable textual verification.}
We define
\begin{equation}
V(q)=\mathbf{1}[\mergecheck\text{ reports a conflict for }h_A,h_B\text{ from }m].
\end{equation}
For fixed Git objects, $V$ is a deterministic checker result for textual integration. A clean $V=0$ still says nothing about compilation or semantics.

\paragraph{Heuristic scheduling.}
A score $s(q)\in\mathbb{R}$ may estimate risk, rank pairs for replay, or delay likely-colliding work. It is never interpreted as a proof. For target conflict coverage $r$, we measure the verifier cost
\begin{equation}
C_s(r)=\min_k\left\{\frac{k}{N}:\frac{\sum_{i=1}^{k}V(q_{\pi_s(i)})}{\sum_j V(q_j)}\ge r\right\},
\end{equation}
where $\pi_s$ orders pairs by decreasing score.

A second requirement is \emph{decision-time discipline}: features used to rank or schedule work must be available before the target checker outcome they are meant to prioritize. T1 therefore uses only open-time information; the lineage model uses authored files and merge-base-to-head history but not the target merge result; and T2 is intentionally conditioned on a replay-confirmed conflict because its task is resolution, not prediction. Table~\ref{tab:layers} summarizes the resulting empirical stack. Counts differ because each layer requires different artifacts and replayability.

\begin{table}[t]
\centering\small
\caption{Empirical layers under one verification contract. Each layer has a distinct target and decision point; counts differ because replayability and metadata requirements differ.}
\label{tab:layers}
\begin{tabularx}{\linewidth}{@{}p{1.7cm}p{2.2cm}Y@{}}
\toprule
Layer & Scale & Grounding and key observation \\
\midrule
Prior population measurement~\citep{xu2026} & 33,596 PRs; 2,807 repos & Published AIDev-pop characterization used as background, not a new estimate in this paper. \\
Stratified replay & 747 sampled pairs; 716 evaluable & Sample-conditioned raw replay: 19.8\% same-agent vs. 41.7\% cross-agent; matched analysis is weaker (Sec.~3.1). \\
Scope corpus & 577,045 pairs; 1,100 repos & 7.38\% share authored files; near-duplicate intent and file scope frequently diverge. \\
Resolution replay & 5,936 successful replays & 1,252 textual conflicts; 23.8\% of locatable pilot hunks exceed a best-of-trivial resolution oracle. \\
Lineage set & 715 complete pairs; 167 conflicts & 47.3\% of conflicts have zero authored overlap; same-information held-out AUROC up to 0.902. \\
Scheduling & 97 repo streams; 19,397 PRs & Scope-aware delaying trades collision exposure against latency/makespan. \\
\bottomrule
\end{tabularx}
\end{table}

\section{Why concurrent coordination needs explicit checks}
\subsection{Concurrency is common; the raw cross-agent gap is not causal}
Prior work on the curated 33,596-PR AIDev-pop subset reports that 26,691 PRs (79.4\%) are open concurrently with another agent PR and that 1,129 of 2,807 repositories (40.2\%) exhibit such activity~\citep{xu2026}. We use those population figures only as background. The 747-pair replay sample used downstream was deliberately stratified: cross-agent pairs are 115/716 evaluable replays (16.1\%), far above their population share, so pooled replay rates are \emph{sample-conditioned}, not population estimates.

In that stratified replay, 48/115 cross-agent pairs conflict textually (41.7\%) versus 119/601 same-agent pairs (19.8\%). A same-repository matched sensitivity materially weakens the interpretation: among 60 repositories with an evaluable pair in both strata, same-agent conflict is 17/60 (28.3\%) and cross-agent conflict 26/60 (43.3\%); the exact McNemar test has $p=0.078$ (15 cross-only versus 6 same-only discordant pairs). A composition check based on agent-specific intra-agent rates expects 37.4\% cross-agent conflict versus the observed 41.7\%. We therefore do \emph{not} claim a cross-vendor penalty. The reliable conclusion for this paper is narrower: concurrent pairwise checking is common enough to motivate an explicit coordination layer, while agent-family effects require larger matched replay.

\subsection{File overlap is informative but not a safety predicate}
The large \mergegym scope corpus contains 577,045 co-active pairs with changed-file lists on both sides. Only 42,600 pairs (7.38\%) share any authored file. Even these overlap events are heterogeneous: 65.0\% share exactly one file and 33.9\% share only configuration, documentation, CI, manifest, or related non-source files. The most common shared basenames include repository-wide artifacts such as \texttt{README.md} and \texttt{package.json}.

Conversely, task-level similarity often appears without file overlap. Among 6,945 near-duplicate-title candidates, 79.6\% share no authored file. Thus path collision and intent collision are not alternate encodings of the same target. A verifiable coordination stack should use each only for the property it actually represents.

\begin{figure}[t]
\centering
\includegraphics[width=0.49\linewidth]{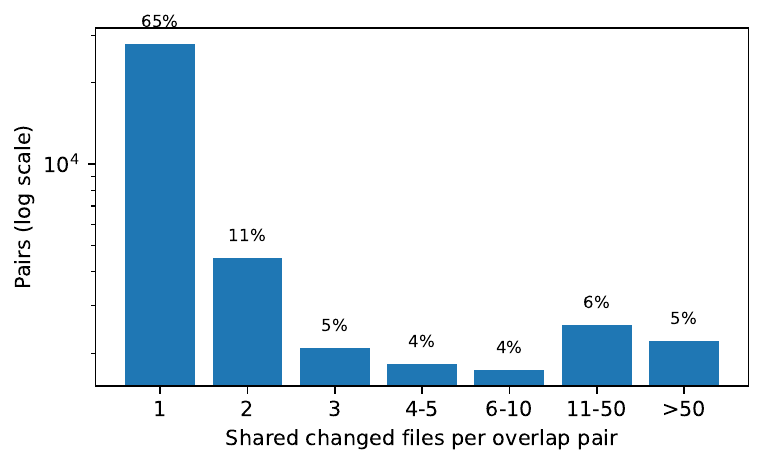}\hfill
\includegraphics[width=0.49\linewidth]{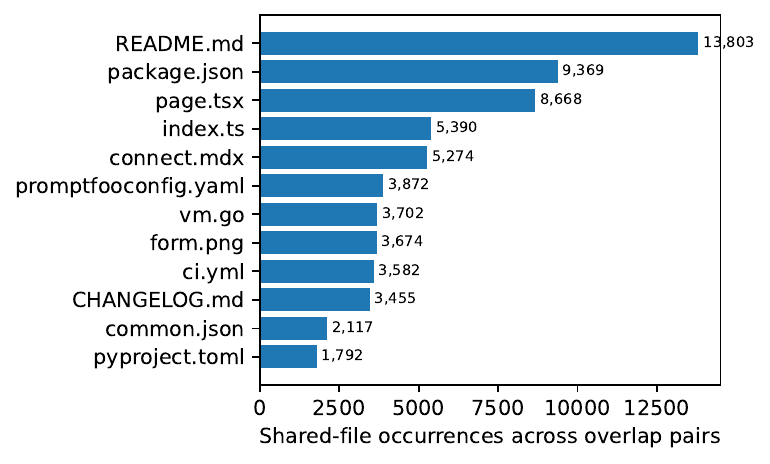}
\caption{\textbf{Scope overlap is a heterogeneous proxy.} Left: most overlapping pairs share one file and one third are pure configuration/documentation overlap. Right: common shared basenames are dominated by repository-wide files.}
\label{fig:scope}
\end{figure}

\section{\mergegym: separate the coordination decisions}
A central failure mode in agent evaluation is target collapse: using a cheap signal as if it were a guarantee, or measuring a system on a label that is unavailable at decision time. \mergegym therefore defines three tracks over a shared pairwise corpus.

\subsection{T1: forecast future implementation scope}
T1 asks whether two open PRs will eventually share an authored file, using only information available at open time: title, body, timing, and optional repository metadata. Training contains 24,283 pairs from 870 repositories; the frozen test set contains 5,387 pairs from 229 repositories, all held out by repository, and is 39.93\% positive after enrichment. The released baseline ladder includes both classical and language-model baselines. Title Jaccard reaches pooled AUC 0.629, TF--IDF cosine 0.644, a repository-held-out logistic model 0.644 pooled / 0.684 within-repository, and a four-feature out-of-fold fusion 0.676. A zero-shot LLM reaches 0.704, and LLM plus all open-time features reaches 0.740 under five-fold out-of-fold fusion. T1 is therefore a useful \emph{planning heuristic}, not a verification result, and these numbers should not be read as saturating the task.

Repository stratification is essential. Pooled merge outcomes exhibit Simpson reversals: for duplicate-title candidates the crude merge odds ratio is 0.851 but the repository-stratified Mantel--Haenszel estimate is 1.507; fix-in-flight candidates similarly reverse from 0.581 to 1.090. A model can otherwise appear to learn interaction risk while mostly learning repository mixture.

\subsection{T2: resolve only executable replay-confirmed conflicts}
T2 changes both the input and the target. From 6,387 both-merged, source-overlap candidates, 5,936 integrations replay successfully and 1,252 (21.1\%) produce textual conflicts under \mergecheck. The benchmark then compares candidate resolutions against the repository's observed reconciliation after both merges, restricted to conflicted hunks.

A 162-conflict core pilot extracts 1,635 hunks, 1,469 of which can be located in the reconciliation. Choosing the later PR side exactly matches 49.6\% of locatable hunks, the earlier side 28.4\%, and union 12.8\%. A best-of-these-trivial-strategies oracle reaches 76.2\%, leaving 23.8\% that require behavior beyond these trivial choices. The individual exact-match percentages sum to 90.8\% because their success sets overlap (for some hunks more than one trivial candidate equals the observed reconciliation); 76.2\% is the union of those success indicators, not their sum. The frozen artifact does not contain a completed learned/LLM T2 resolver run, so we do not claim that the trivial floor is competitive with modern resolvers. This floor matters only as a reproducible minimum: ``the output parses'' or ``the conflict markers are gone'' is not enough to establish meaningful synthesis.

\subsection{T3: schedule under an explicit cost function}
T3 replays 97 repository streams (19,397 PRs; 38,526 observed scope-collision pairs). Each episode contains arrivals, durations, open-time features, and the observed pairwise scope relation. An online policy decides START or DELAY; 778 right-censored durations are capped under the replay protocol. The objective is deliberately operational: reduce observed scope-collision exposure, net of newly created overlaps, while reporting added delay or makespan inflation rather than claiming semantic safety.

FIFO serialization removes all observed scope collisions but at 167.3\% median makespan inflation (repository-bootstrap 95\% CI for the median: 114.2--195.7\%). An oracle gate with perfect knowledge of future scope overlap also reaches 100\% de-overlap at 33.4\% median inflation (20.0--47.4\%). A trained gate using only decision-time features (held-out AUC 0.66 in the released artifact) traces the realizable frontier; at threshold 0.20 it achieves 91.7\% median de-overlap (81.2--97.5\%) at 65.0\% median inflation (35.7--99.2\%). These bootstrap intervals resample the 97 repository episodes. Excluding the two mega-repositories leaves the medians essentially unchanged. These are scheduling-feasibility results under frozen labels and non-reactive agents, not claims that textual or semantic conflicts were prevented.

\begin{figure}[t]
\centering
\includegraphics[width=0.49\linewidth]{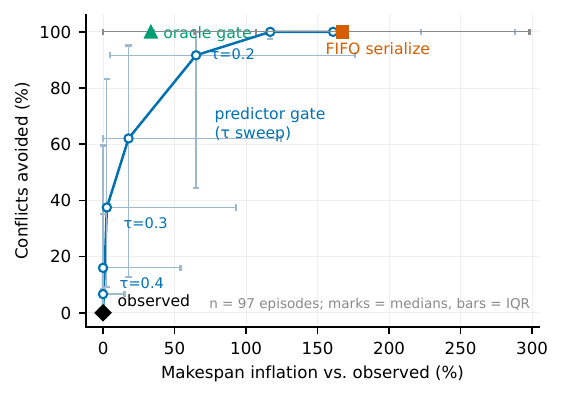}\hfill
\includegraphics[width=0.49\linewidth]{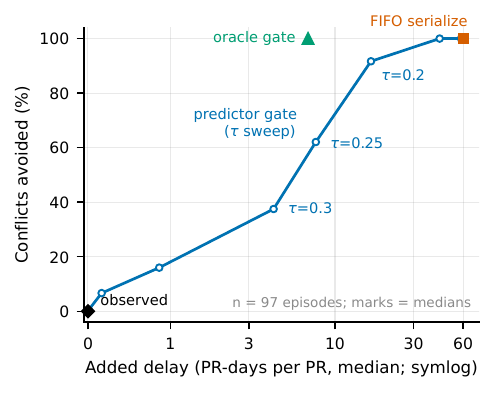}
\caption{\textbf{Scheduling is a cost--benefit problem, not a proof.} Across 97 replayed repository streams, scope-aware gating removes observed collision exposure much more efficiently than full FIFO serialization.}
\label{fig:scheduling}
\end{figure}

\section{Authored-diff proxy mismatch and branch lineage}
The benchmark decomposition identifies when exact replay is required, but a high-throughput coordinator may still want to prioritize checks. The central observation here is not a newly discovered Git mechanism. A three-way merge compares both heads relative to their two-head merge base, while our cheap authored-file proxy compares each PR with its own PR base. Those are different diffs. We measure how often that mismatch matters and whether decision-time lineage metadata recovers useful ranking signal.

\subsection{The zero-overlap stratum}
The lineage evaluation starts from the stratified replay sample: 716 pairs have executable labels, and 715 also have both authored file lists. It contains 600 same-agent pairs (119 conflicts) and 115 cross-agent pairs (48 conflicts), for 167 conflicts and 548 clean pairs overall. Because cross-agent pairs are intentionally enriched, neither prevalence nor AUROC on this set should be interpreted as a population quantity. Of the 167 conflicts, 88 have authored-file overlap and \textbf{79 do not}. Any detector that \emph{requires} $D_A\cap D_B\neq\varnothing$ before predicting conflict therefore has maximum recall 88/167 = 52.7\% on this evaluation set.

The zero-overlap cases are consistent with the expected diff mismatch. For 64/79 (81.0\%) zero-overlap conflicts, every conflicted path appears in exactly one PR's authored diff; only nine pairs have all conflicted paths absent from both authored diffs. Across zero-overlap conflicted paths, 74.8\% use source-code extensions and only 2.7\% are lockfiles. A file-list-only Claude Haiku 4.5 baseline, which sees repository identifier and authored paths but no lineage, recalls 92.0\% of overlap-visible conflicts and only 2/79 (2.5\%) of zero-overlap conflicts. This is a result for that restricted information set, not a claim about language models in general.

\begin{figure}[t]
\centering
\includegraphics[width=\linewidth]{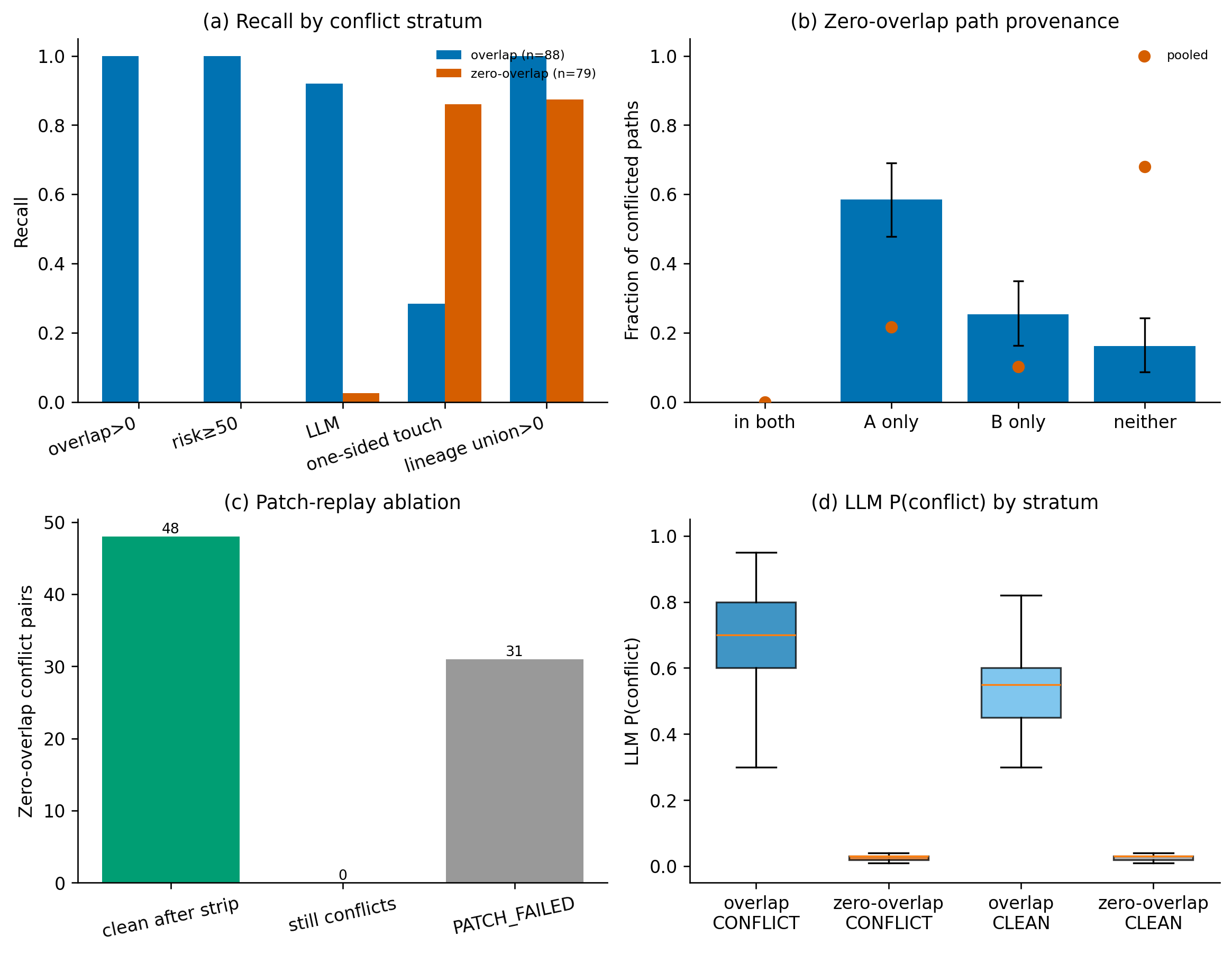}
\caption{\textbf{Authored diffs omit lineage state used by a head-to-head merge.} (a) Recall by overlap stratum. (b) Conflicted-path provenance in zero-overlap cases. (c) Patch reconstruction is successful for 48 cases and all become clean; 31 reconstruction failures are inconclusive, not negative results. (d) The evaluated file-list LLM assigns low conflict probability to the zero-overlap stratum.}
\label{fig:mechanism}
\end{figure}

\subsection{Decision-time lineage features and stronger baselines}
Let $L_A,L_B$ denote paths changed on the common-base-to-head histories $m\rightarrow h_A$ and $m\rightarrow h_B$. We define
\begin{align}
T(A,B)=\mathbf{1}[&(D_A\setminus D_B)\cap L_B\neq\varnothing \\
&\lor (D_B\setminus D_A)\cap L_A\neq\varnothing].
\end{align}
The Boolean \touch feature asks whether the counterparty lineage touched a path authored only on the other side. On zero-overlap conflict pairs it fires on 68/79 (86.1\%); on zero-overlap clean pairs it fires on 17.1\%. A still simpler standalone rule---whether the two common-base-to-head lineage unions touch any common path---already reaches held-out AUROC 0.877 and recall 0.949 at its binary operating point. This addresses the concern that the original paper compared the 12-feature model against unnecessarily weak lineage baselines.

We fit \linmodel using 12 decision-time features: authored overlap count, Jaccard overlap, directory overlap, deterministic metadata risk, \touch, lineage-union overlap, two behind-base indicators, two base-churn counts, and two branch ages. Repositories are deterministically hashed into dev (218 pairs, 50 conflicts) and held-out test (497 pairs, 117 conflicts). Numeric features are standardized on dev and the model is fit on dev only. We additionally report an untuned random forest (500 trees, default split/leaf settings, seed 0) on the identical 12 features; no test-set tuning is performed.

\begin{table}[t]
\centering\small
\caption{Repository-held-out test performance on 497 pairs (117 conflicts). AUROC intervals are 95\% bootstrap CIs where shown. PR-AUC is average precision. The random forest is a stronger same-information tabular baseline.}
\label{tab:lineage}
\begin{tabular}{lcc}
\toprule
Method & AUROC & PR-AUC \\
\midrule
File overlap & 0.694 [0.646, 0.743] & 0.424 \\
Metadata risk & 0.743 [0.692, 0.791] & 0.428 \\
File-list LLM & 0.732 [0.673, 0.787] & 0.555 \\
\touch & 0.717 [0.668, 0.765] & 0.417 \\
Lineage-union rule & 0.877 [0.842, 0.910] & 0.611 \\
\linmodel & 0.882 [0.845, 0.917] & 0.661 \\
Random forest, same 12 features & \textbf{0.902 [0.875, 0.928]} & \textbf{0.701} \\
\bottomrule
\end{tabular}
\end{table}

The stronger tree baseline is an important qualification: the transparent logistic model is not close to an established upper bound. The lineage-union rule nearly matches its AUROC, while the random forest improves both AUROC and PR-AUC. The feature-subset ablations and alternative repository splits in Appendix~\ref{app:robustness} show that most of the gain comes from lineage state rather than authored-overlap features alone.

\subsection{Controlled reconstruction check}
For each of the 79 zero-overlap conflicts we attempt to reconstruct both PR-authored patches on clean branches rooted at the two-head merge base and rerun \mergecheck. Forty-eight pairs are reconstructable and all 48 become clean; none remain conflicting. The remaining 31 are inconclusive: 12 patches conflict while being applied to the reconstruction base, 7 reference paths absent from that base, 1 hits binary-patch/index limitations, 2 fail during clone/network retrieval, and 9 have other patch-application failures. Because disjoint authored patches reconstructed on a common base are expected to merge cleanly, this experiment is best interpreted as a \emph{pipeline validation of the proxy-mismatch explanation}, not a discovery of a hidden causal mechanism.

A separate question remains unresolved: whether each PR would already conflict with the contemporaneous default-branch tip, independent of the other PR. The frozen lineage package does not retain a time-indexed default-branch tip for every pair, so we cannot compute that baseline retrospectively without a new historical-data collection. We therefore do not claim that all 79 cases are uniquely pairwise rather than partly reflecting branch staleness.

\section{Verifier-guided triage on the held-out test}
Operationally, AUROC is secondary to how many failures a fixed replay budget surfaces. On the 497-pair held-out test, \linmodel recovers 32.5\%, 58.1\%, 81.2\%, and 96.6\% of the 117 conflicts after replaying the top 10\%, 20\%, 33.3\%, and 50\% of pairs, respectively; precision at those budgets is 76.0\%, 68.0\%, 57.2\%, and 45.4\%. The untuned random forest recovers 32.5\%, 57.3\%, 82.9\%, and 96.6\% at the same budgets. The held-out logistic model has Brier score 0.113 [0.096, 0.132], so the evaluation reports calibration as well as ranking.

\begin{figure}[t]
\centering
\includegraphics[width=\linewidth]{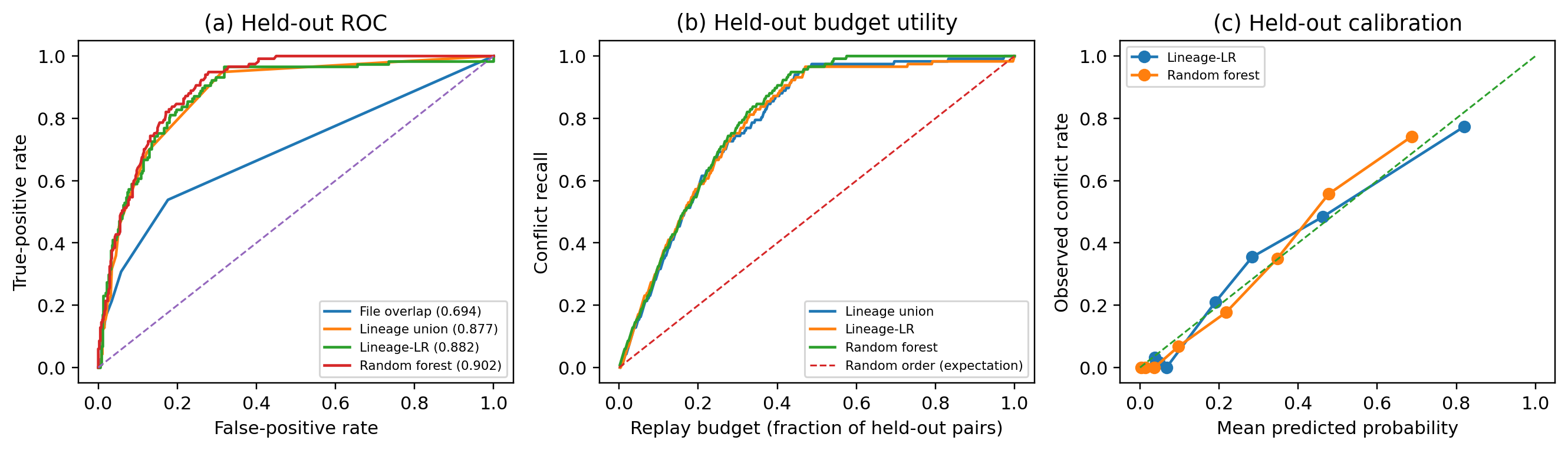}
\caption{\textbf{Held-out verifier prioritization.} (a) ROC on the 497-pair repository-held-out test. (b) Conflict recall as a function of exact-replay budget. (c) Quantile-binned calibration for the two learned models. The checker remains authoritative; curves only prioritize its use.}
\label{fig:evaluation}
\end{figure}

The cost interpretation is deliberately narrow. The released T2 replay logs contain 5,940 timing records: local three-way replay has median runtime 0.02 s, 90th percentile 0.38 s, and 95th percentile 0.50 s. Lineage feature collection itself can require remote compare metadata, and the original crawl did not record end-to-end feature-fetch timing. Therefore this paper does \emph{not} demonstrate that predicting before \mergecheck is cheaper than simply running \mergecheck. If local textual replay is available and cheap, verify every pair and discard the predictor. Triage becomes relevant only when the authoritative chain is costlier (e.g., checkout/build/test/specification checks), rate-limited, remote, or queued behind other verifier work.

\section{Design implication beyond textual checking}
The experiments concern only textual merge replay. We have not evaluated build, test, semantic, or specification verifiers in this paper. The following is therefore a design implication rather than an empirical result: the same trusted-computing-base discipline can be extended to stronger predicates. A concurrent coding stack can expose a sequence of increasingly strong predicates:
\begin{equation}
\phi_{text}\rightarrow\phi_{build}\rightarrow\phi_{test}\rightarrow\phi_{spec},
\end{equation}
where the corresponding backends might be Git merge replay, a compiler/static analyzer, an executable test suite, and a model checker, SMT solver, proof assistant, or equivalence checker. Passing an earlier predicate is never silently promoted into evidence for a later one. Learned models can propose repairs, invariants, schedules, or promising proof obligations, but each claim remains attached to the backend that checks it. This makes the trust boundary auditable: a failure of the heuristic can waste verifier effort or latency, but it cannot by itself upgrade an unchecked pair into a certified one.

This perspective also clarifies human alignment. A maintainer should be able to see \emph{why} work was delayed or escalated: e.g., ``authored files are disjoint, but the counterparty lineage modified a one-sided path; exact replay is queued.'' Such explanations calibrate trust better than a generic ``87\% conflict probability,'' because they expose both the heuristic evidence and the still-unchecked property.

The same separation helps benchmark design. Pass@k is insufficient when the relevant question is whether the system eventually establishes an explicit guarantee, how much checker compute it spends, and which failures remain outside the trusted computing base. Our held-out replay-budget curves and scheduling frontiers are narrow examples of such metrics. Extending the evaluation to build/test/specification backends is future work.

\section{Related work}
Classical software-merging research distinguishes textual, syntactic, build, and semantic interference~\citep{mens2002,horwitz1989}. Palant\'ir and speculative-merging systems surface collaboration conflicts before integration~\citep{sarma2012,brun2011}; Cassandra schedules tasks to reduce expected conflict~\citep{kasi2013}. Static semantic-conflict detection and verified three-way merge move beyond textual checking toward program behavior~\citep{santos2023,sousa2018}. ConflictBench evaluates merge tools on curated scenarios~\citep{shen2024conflictbench}.

Agentic software engineering introduces a different throughput regime. AIDev provides a large observational substrate for coding-agent activity~\citep{li2026aidev}; AgenticFlict measures conflicts involving agent-authored PRs~\citep{agenticflict}; SWE-bench evaluates isolated issue resolution against executable tests~\citep{jimenez2024}. CooperBench and BulkPR-Bench move toward interacting or queue-level agent evaluation~\citep{cooperbench,bulkpr}. Our contribution is complementary: we connect in-the-wild pairwise measurement to an explicit verifier-in-the-loop contract, distinguish observable scope from executable conflict, and show that asymmetric repository lineage materially improves which exact checks should run first.

\section{Limitations and responsible claims}
\textbf{Textual property only.} \mergecheck establishes Git's textual merge outcome for fixed heads and a merge base. CLEAN does not imply build success, passing tests, semantic compatibility, security, or alignment with user intent. Section~7 is a design implication, not an evaluated stronger-verifier result.

\textbf{Small, stratified lineage evaluation.} The primary predictive test has 497 pairs and 117 conflicts. The 715-pair lineage set intentionally over-samples cross-agent pairs and is not population-representative. AUROC/PR-AUC are therefore held-out performance on this evaluation distribution, not prevalence or deployment-value estimates.

\textbf{Model headroom remains.} A simple lineage-union rule reaches AUROC 0.877 and an untuned random forest reaches 0.902, exceeding the transparent logistic model's 0.882. We therefore claim that lineage state is useful, not that \linmodel is close to the strongest achievable predictor under the same information constraints.

\textbf{Repository, language, and agent concentration.} Thirty alternative 30/70 repository-held-out splits give median full-model AUROC 0.876 (IQR 0.865--0.885; range 0.829--0.901). Leave-one-agent-family-out AUROCs range 0.845--0.901 across the five represented agent families; leave-one-language-out AUROCs over the six languages with at least 40 pairs range 0.835--0.950. These checks reduce but do not remove external-validity concerns, especially for repositories and agent families absent from AIDev-pop.

\textbf{Lineage availability.} One of 715 pairs lacks lineage retrieval and 60/714 retrieved rows are marked truncated by the compare-data path; 44 truncated-or-missing rows fall in the held-out test. On the 453 untruncated test rows, AUROC is 0.904. A conservative stress test that zeros lineage features for truncated/missing test rows lowers full-model AUROC from 0.882 to 0.791 and PR-AUC from 0.661 to 0.574. A deployment should therefore fall back to exact verification when lineage metadata is incomplete rather than treating missing state as evidence of safety.

\textbf{Pairwise versus staleness.} The frozen data do not preserve the contemporaneous default-branch tip needed to test whether each PR independently conflicts with mainline. The zero-overlap result is thus framed as a head-to-head proxy mismatch, not proof that every case is uniquely caused by interaction between the two PRs.

\textbf{Scheduling replay.} \mergegym T3 freezes historical labels, keeps durations fixed, and assumes agents do not react to delay. Its frontier is a scheduling-feasibility analysis, not a causal deployment trial. Bootstrap intervals quantify episode heterogeneity but do not repair this counterfactual limitation.

\textbf{Verifier cost.} Local textual replay is already very cheap in the released timing logs, and lineage-fetch cost was not captured consistently. We make no net-compute-savings claim for \mergecheck triage alone.

\textbf{Code provenance and licensing.} Source-repository licenses govern code originating from those repositories. The released supplement excludes the raw T2 hunk-text bundle pending repository-by-repository redistribution clearance and retains only derived labels/metrics plus scripts and identifiers needed to reconstruct hunks from upstream sources where permitted. This licensing constraint is distinct from scientific reproducibility and remains explicit.

\textbf{Potential impact and misuse.} Low heuristic scores must never be treated as permission to skip a checker whose property is required. The mitigation is architectural: scores may prioritize or delay checks, but guarantees come only from the corresponding executable backend.

\section{Conclusion}
Concurrent coding agents create a verification-orchestration problem: cheap scope proxies can be useful for planning, but they are not interchangeable with executable checks. In the stratified lineage set, 47.3\% of replay-confirmed textual conflicts occur despite disjoint authored file sets because authored PR diffs and the diff used by a two-head three-way merge are different objects. The contribution is the measured prevalence and operational consequence of that proxy mismatch, not a new Git mechanism.

Decision-time lineage state substantially improves held-out prioritization. A standalone lineage-union rule reaches AUROC 0.877, the transparent logistic model 0.882 with PR-AUC 0.661, and an untuned random forest on the same 12 features 0.902 with PR-AUC 0.701. At a 33.3\% replay budget the learned models recover roughly 81--83\% of held-out conflicts. The correct systems principle remains: \emph{heuristics should schedule guarantees, not impersonate them}. When the checker is cheap, run it on everything; when a stronger verifier chain is expensive or constrained, use predictions only to order work while leaving the executable backend in charge of the claim.

\appendix
\section{Robustness analyses}\label{app:robustness}
All analyses in this appendix use the frozen released row-level artifact; no outcome is imputed and no test labels are used for model fitting.

\paragraph{Claim taxonomy.}
Population, replay-rate, zero-overlap, and patch-reconstruction numbers are descriptive characterizations of their stated samples. T1 and lineage AUROC/PR-AUC/budget results are predictive evaluations on repository-held-out data. T3 numbers are frozen-label scheduling simulations. None of the descriptive rates are presented as population prevalence unless explicitly inherited from the prior AIDev-pop measurement in Table~\ref{tab:layers}.

\paragraph{Feature-subset ablation.}
We refit logistic regression on the same deterministic repository split using nested decision-time feature groups. ``Overlap-only'' uses overlap count, Jaccard, and directory overlap; ``metadata-only'' adds the deterministic risk score; ``lineage-only'' uses \touch, lineage-union overlap, two behind-base indicators, two base-churn counts, and two branch ages; ``full'' is the 12-feature model.
\begin{table}[h]
\centering\small
\caption{Held-out feature ablation on 497 pairs. Brier is lower-is-better.}
\begin{tabular}{lrrr}
\toprule
Feature set & AUROC & PR-AUC & Brier \\
\midrule
Overlap-only & 0.702 & 0.444 & 0.159 \\
Metadata-only & 0.748 & 0.482 & 0.154 \\
\touch only & 0.717 & 0.417 & 0.150 \\
Lineage-only & 0.873 & 0.636 & 0.128 \\
Full 12-feature LR & 0.882 & 0.661 & 0.113 \\
\bottomrule
\end{tabular}
\end{table}

\paragraph{Replay-budget utility.}
For \linmodel, precision/recall at budgets 10/20/33.3/50\% are 0.760/0.325, 0.680/0.581, 0.572/0.812, and 0.454/0.966. For the untuned random forest they are 0.760/0.325, 0.670/0.573, 0.584/0.829, and 0.454/0.966. These are computed only on the held-out test and are the central operational result.
At the original dev-selected binary threshold, \linmodel recall is 49.6\%, below file-overlap recall 53.8\%; the useful result is therefore ranking/budget utility rather than that single operating point.

\paragraph{Alternative repository splits.}
Across 30 GroupShuffleSplit 30/70 repository-held-out partitions (random state sequence fixed from 2026), the full logistic model has median AUROC 0.876, IQR 0.865--0.885, range 0.829--0.901; median PR-AUC 0.639, IQR 0.616--0.665. The lineage-only feature subset has median AUROC 0.805 (IQR 0.784--0.850), showing more split sensitivity than the full model but preserving substantial signal.

\paragraph{Agent- and language-group sensitivity.}
On the deterministic held-out test, full-model AUROC is 0.885 on same-agent pairs ($n=414$) and 0.852 on cross-agent pairs ($n=83$). Leave-one-agent-family-out AUROCs are 0.845 (Claude Code), 0.860 (Copilot), 0.845 (Cursor), 0.850 (Devin), and 0.901 (OpenAI Codex). Joining repository primary language from the frozen repository table, leave-one-language-out AUROCs for the six languages with at least 40 lineage pairs are 0.887 (TypeScript), 0.891 (Python), 0.860 (Go), 0.900 (C\#), 0.950 (Rust), and 0.835 (JavaScript). These are robustness checks, not claims of uniform cross-domain generalization.

\paragraph{Missing/truncated lineage.}
Lineage is missing for 1/715 pairs; 60/714 retrieved rows carry a truncation flag, including 44 held-out rows. Evaluating only the 453 held-out rows without missing/truncated lineage gives AUROC 0.904. Conservatively zeroing all eight lineage features for truncated/missing held-out rows reduces full-model AUROC to 0.791 and PR-AUC to 0.574. This motivates an explicit deployment fallback to the exact checker.

\paragraph{Why \touch misses 11 zero-overlap conflicts.}
Eight of the 11 \touch false negatives carry truncated lineage metadata. Under the retrieved metadata, 10/11 also have zero lineage-union overlap; the remaining case has lineage-union overlap but no one-sided authored-path intersection and contains multiple conflict types. None of the 11 is labeled as a rename conflict in the replay output. Thus the misses are dominated by unavailable/incomplete lineage state or by conflict structure not captured by the one-sided authored-path predicate, rather than a single rename pattern.

\paragraph{T3 episode uncertainty.}
Repository-bootstrap 95\% intervals for medians over the 97 episodes are: FIFO makespan inflation 167.3\% [114.2, 195.7]; oracle inflation 33.4\% [20.0, 47.4]; predictor($\tau=0.20$) de-overlap 91.7\% [81.2, 97.5] and makespan inflation 65.0\% [35.7, 99.2]. All three policies use the same frozen-label simulator assumptions.

\paragraph{Ablation failures.}
The 31 inconclusive patch reconstructions comprise 12 patch conflicts on the reconstruction base, 7 paths absent from the reconstruction index, 1 binary-patch/index limitation, 2 clone/network failures, and 9 other patch-application failures. They are never counted as clean or still-conflicting outcomes.

\paragraph{T1 and T2 baseline scope.}
The full T1 leaderboard in the artifact includes nine baselines: repository base rate 0.601 AUC, overlap-hours 0.570, filename-token overlap 0.574, title Jaccard 0.629, TF--IDF cosine 0.644, repository-held-out LR 0.644, four-feature OOF fusion 0.676, zero-shot LLM 0.704, and LLM+all-feature OOF fusion 0.740. The T2 artifact has no completed learned/LLM resolver result. The trivial resolver percentages overlap by hunk, explaining why 49.6+28.4+12.8=90.8\% while the union-of-successes oracle is 76.2\%. At the original dev-selected threshold, the metadata-risk binary decisions coincide with the overlap rule; its value is as a continuous ranking baseline, not as a distinct fixed decision rule.

\paragraph{Verifier timing.}
The T2 replay table contains 5,940 timing rows: median 0.02 s, 90th percentile 0.38 s, and 95th percentile 0.50 s per local replay. End-to-end lineage-fetch timing was not recorded. Accordingly, no claim of net savings from lineage triage over local \mergecheck is made.

\section{Additional benchmark details}
\paragraph{T1 baseline ladder.}
On the 5,387-pair frozen T1 test set, overlap-hours AUC is 0.570; filename-token overlap 0.574; repository base rate 0.601; title Jaccard 0.629; TF--IDF cosine 0.644; repository-held-out logistic regression 0.644 pooled and 0.684 within-repository; four-feature out-of-fold fusion 0.676; zero-shot LLM 0.704; and LLM-plus-all-feature out-of-fold fusion 0.740.

\paragraph{T2 replay composition.}
Conflict yield rises with PR size (10.3\% for pairs at $\le 1$k modified lines to 34.3\% above 20k) and shared source-file count (5.0\% with one shared source file to 24.0\% with six or more). Cross-agent candidates conflict at 31.8\% versus 20.9\% for same-agent candidates within this filtered replay pool.

\paragraph{Lineage evaluation protocol.}
Repositories are deterministically hashed into dev (218 pairs, 50 conflicts) and test (497 pairs, 117 conflicts). Numeric lineage-model features are standardized on dev; model fitting and operating-threshold selection use dev only. Test AUROC confidence intervals use 2,000 bootstrap resamples with seed 0. The file-list LLM baseline is Claude Haiku 4.5; its prompt contains the repository identifier, each PR's authored paths (capped at 150 per side with truncation marked), and a request for a conflict verdict, confidence, and short rationale. It receives no merge result, conflicted path, conflict type, or lineage variable.

\section{Artifact and exact reproduction map}
\paragraph{Artifact.}
The frozen review artifact is available at \url{https://doi.org/10.17605/OSF.IO/38SKH}. It unifies the three constituent research packages used by this paper. The release is intentionally layered so that readers can verify headline tables from frozen artifacts without repeating network-dependent GitHub fetches, while still retaining the scripts needed to rerun those live stages when upstream refs remain available.

\begin{table}[h]
\centering\small
\caption{Artifact map. The OSF bundle preserves frozen inputs/results and the scripts that produced them; network-dependent replay stages are separated from offline regeneration.}
\label{tab:artifact-map}
\begin{tabularx}{\linewidth}{@{}p{2.1cm}p{3.0cm}Y@{}}
\toprule
Component & Frozen artifacts & Primary reproduction path \\
\midrule
Population/\\replay & AIDev-derived extracts, legacy 747-pair replay tables, environment record, analysis plan, claim-to-script map & \texttt{make data analysis figures}; live confirmation via \texttt{make replay} when Git refs remain available \\
\mergegym & 577,045-pair corpus, T1 train/test manifests and score vectors, T2 replay/hunk tables, T3 episode results and summaries & T1 scripts in order; \texttt{t2\_groundtruth.py} over released replay output; \texttt{t3\_simulator.py --check} for observed/FIFO/oracle episodes \\
Lineage & 715 labeled pairs, 714/715 lineage-feature records, verified Haiku predictions, patch-stripping outcomes, generated numbers/figures & \texttt{python metrics.py}; \texttt{python figures.py}; optional live reruns for lineage fetch, LLM calls, and failed ablations \\
\bottomrule
\end{tabularx}
\end{table}

\paragraph{Population and replay provenance.}
The population/replay component records the upstream AIDev snapshot hashes and the exact Git configuration used by replay. The recorded source files have SHA-256 hashes \texttt{c0b8e81e...f3e8} for \texttt{pull\_request.parquet} and \texttt{a08e34be...6aa7} for \texttt{repository.parquet}; the artifact verifies those hashes before analysis. Replay forces \texttt{merge.conflictStyle=merge}, \texttt{diff.renameLimit=400}, and rename detection. Its committed reproduction notes distinguish offline regeneration from live GitHub fetches and explicitly treat deleted/private refs as unavailable rather than silently replacing pairs.

\paragraph{\mergegym manifests.}
The T1 manifest freezes seed 0, 24,283 training pairs from 870 repositories, and 5,387 test pairs from 229 repositories, with every test repository unseen in training. The manifest also records the 200-pair-per-repository training cap, mega-repository exclusions used for the training pool, and the train-only feature construction. T2 releases all 6,387 candidates, replay status for 5,936 successful integrations, 1,252 conflict labels, and the hunk/reconciliation tables used for the trivial-resolution floor. T3 releases all 97 episode rows plus a summary over 19,397 PRs and 38,526 labeled scope-collision pairs; \texttt{--check} verifies the observed, FIFO, and oracle episode mechanics against the frozen result table.

\paragraph{Lineage package and generated numbers.}
The lineage component releases \texttt{benchmark\_pairs.jsonl}, \texttt{lineage\_features.jsonl}, \texttt{predictions\_haiku.jsonl}, \texttt{ablation\_results.csv}, and the scripts that regenerate \texttt{numbers.json}, \texttt{paper/numbers.tex}, and every lineage figure. The LLM runner requires an explicit model argument and records the token-consuming served model; the released baseline verifies \texttt{claude-haiku-4-5-20251001}. Rows with a requested/served model mismatch are rejected rather than relabeled. The main held-out AUROC, zero-overlap counts, \touch recall, DeLong comparisons, and coverage--cost operating points are therefore generated from committed row-level artifacts rather than typed into the manuscript.

\paragraph{Frozen versus live reproduction.}
Some operations are intrinsically time-sensitive. Replaying Git heads requires public refs that may later disappear, and rerunning the external LLM baseline requires model access. The artifact therefore ships the frozen replay/LLM outputs used by the paper together with their manifests and verification scripts. This makes the reported analysis reproducible even if an upstream PR ref or hosted model later becomes unavailable, while still providing the live-fetch code for independent replication. T2 records 447 unavailable pairs and four missing refs in the full candidate set instead of imputing outcomes. The lineage package similarly keeps 31 failed patch reconstructions as inconclusive rather than folding them into either ablation result.

\paragraph{Software environment and compute.}
The released code is CPU-oriented: Git replay, Pandas/NumPy/SciPy/statistical routines, scikit-learn-style linear models, and plotting; no GPU training is required for the proposed method. The population package provides a Dockerfile, pinned Python 3.12.8 image path for its containerized workflow, a recorded local Git/Python environment, and stage-level runtime guidance (offline analysis and figures are minute-scale; live replay is hours-scale). The \mergegym artifact provides a Python dependency lockfile and the lineage package provides its minimal requirements. Historical peak memory and wall-clock accounting were not captured consistently across all three original development environments, so we make no precise peak-memory or total-compute claim on that point.

\paragraph{Asset provenance and licensing.}
Upstream AIDev packaging is documented as CC-BY-4.0; source-repository licenses govern repository code. The released supplement therefore omits the raw T2 hunk-text bundle and retains derived labels/metrics, provenance identifiers, and reconstruction scripts. Excerpts can be reconstructed from upstream sources when permitted by their licenses; no scientific claim depends on redistributing them.

\end{document}